\documentclass[pre,showpacs,preprint,tightenlines,nofootinbib]{revtex4}

\usepackage{graphicx,bm,amsfonts}
\usepackage[tbtags]{amsmath}

\newcommand{\ij}{i\kern -0.08em j}
\newcommand{\half}{{\textstyle\frac{1}{2}}}
\def\hn{\mskip-0.5\thinmuskip}
\def\hp{\mskip0.5\thinmuskip}

\def\diag{\mathop{\rm diag}\nolimits}
\def\beq{\begin{equation}}
\def\eeq{\end{equation}}
\def\eeql#1{\label{#1} \end{equation}}

\newcommand{\To}{\rightarrow}

\newcommand{\pt}{\partial}

\newcommand{\Oc}{\mathcal{O}}
\newcommand{\Ec}{\mathcal{E}}

\newcommand{\pht}{{\vphantom{2}}}

\begin{document}

\title{Mediated tunable coupling of flux qubits}
\author{Alec \surname{Maassen van den Brink}}
\email{alec@dwavesys.com}
\affiliation{D-Wave Systems Inc., 100-4401 Still Creek Drive, Burnaby, B.C., V5C 6G9 Canada}
\author{A.J. Berkley}
\email{ajb@dwavesys.com}
\affiliation{D-Wave Systems Inc., 100-4401 Still Creek Drive, Burnaby, B.C., V5C 6G9 Canada}
\author{M.~Yalowsky}
\email{yalowsky@interchange.ubc.ca}
\affiliation{D-Wave Systems Inc., 100-4401 Still Creek Drive, Burnaby, B.C., V5C 6G9 Canada}
\date{\today}

\begin{abstract}
It is sketched how a monostable rf- or dc-SQUID can mediate an inductive coupling between two adjacent flux qubits. The nontrivial dependence of the SQUID's susceptibility on external flux makes it possible to continuously tune the induced coupling from antiferromagnetic (AF) to ferromagnetic (FM). In particular, for suitable parameters, the induced FM coupling can be sufficiently large to overcome any possible direct AF inductive coupling between the qubits.

The main features follow from a classical analysis of the multi-qubit potential. A fully quantum treatment yields similar results, but with a modified expression for the SQUID susceptibility.

Since the latter is exact, it can also be used to evaluate the susceptibility---or, equivalently, energy-level curvature---of an isolated rf-SQUID for larger shielding and at degenerate flux bias, i.e., a (bistable) qubit. The result is compared to the standard two-level (pseudospin) treatment of the anticrossing, and the ensuing conclusions are verified numerically.

\end{abstract}

%Quantum computing
\maketitle

\section{Introduction}

Superconducting devices made of mesoscopic Josephson junctions are one of the few remaining viable candidates for the realization of quantum bits (qubits). The major types encode quantum information in the charge, phase, or flux degree of freedom~\cite{mss}. Quantum coherence in such devices was conclusively demonstrated at the turn of the century. While much remains to be done in improving the coherence times, control, and readout of individual qubits, another key challenge on the road towards quantum-register prototypes is the coupling of several qubits. In the past years, coupling was achieved for each of the mentioned qubit types~\cite{coupl-ref,IMT2qb}.

In all published experiments, the coupling strength is a constant determined at the time of fabrication. For each model of quantum computing (gate-operations, adiabatic/ground-state, etc.), however, it would be very desirable to have the coupling tunable \emph{in situ}, even if this requirement can sometimes be circumvented in ``software"~\cite{feldman}. In particular, whenever the coupling can be continuously tuned between two values with opposite signs, it can be switched off naturally.

For charge qubits, a design fulfilling these requirements was presented in Ref.~\cite{AB}. The coupling between the two qubits is \emph{mediated} via a third device, the bias parameters of which can be used to tune the coupling constant. While the coupling element thus is nontrivial, it is important that it be passive, i.e., that it does not get entangled with the qubits so that the system's total effective Hilbert space remains four-dimensional. In practice, this means that the coupler must stay in its ground state, adiabatically following the qubit dynamics.

Mediated coupling can be adapted to flux qubits~\cite{Plourde}. This paper's purpose is to contribute to its analytical theory. The main results can be obtained from a purely classical analysis of the system's potential, which may have some independent pedagogical interest. This is carried out in Sec.~\ref{class} for two choices of the coupler, both SQUID devices. A full quantum analysis is presented in Sec.~\ref{quant}; the main result for the tunable coupling constant is highly analogous to its classical counterpart, but some additional effects are uncovered concerning a renormalization of the qubit parameters. In the formula for the coupling constant, the nontrivial factor is an \emph{exact} expression for the coupler's ground-state magnetic susceptibility. In Sec.~\ref{bist}, this is applied to a different regime, where the coupler behaves like a qubit itself, allowing for a comparison with the predictions of the two-state model. Some concluding remarks are made in Sec.~\ref{discuss}.

\section{Classical analysis}
\label{class}

\subsection{rf-SQUID coupler}
\label{rf}

\begin{figure}[th]
\begin{picture}(106,33)
  \put(0,0){\line(1,0){30}}
  \put(0,0){\line(0,1){30}}
  \put(0,30){\line(1,0){30}}
  \put(30,0){\line(0,1){30}}
  \put(12,27){\line(1,1){6}}
  \put(12,33){\line(1,-1){6}}
  \put(13,14){$\phi_a^\mathrm{x}$}
  \put(13,24){$E_a$}
  \put(2,2){$L_a$}
  \put(30.7,14){$M_{ab}$}
  \put(38,5){\line(1,0){30}}
  \put(38,5){\line(0,1){20}}
  \put(38,25){\line(1,0){30}}
  \put(68,5){\line(0,1){20}}
  \put(50,22){\line(1,1){6}}
  \put(50,28){\line(1,-1){6}}
  \put(51,14){$\phi_b^\mathrm{x}$}
  \put(51,29){$E_b$}
  \put(40,7){$L_b$}
  \put(68.8,14){$M_{bc}$}
  \put(76,0){\line(1,0){30}}
  \put(76,0){\line(0,1){30}}
  \put(76,30){\line(1,0){30}}
  \put(106,0){\line(0,1){30}}
  \put(87,27){\line(1,1){6}}
  \put(87,33){\line(1,-1){6}}
  \put(88,14){$\phi_c^\mathrm{x}$}
  \put(88,24){$E_c$}
  \put(77,2){$L_c$}
\end{picture}
\caption{A system of three rf-SQUIDs, of which the two outer ones function as qubits and the inner one as a coupler.}
\label{fig1}
\end{figure}
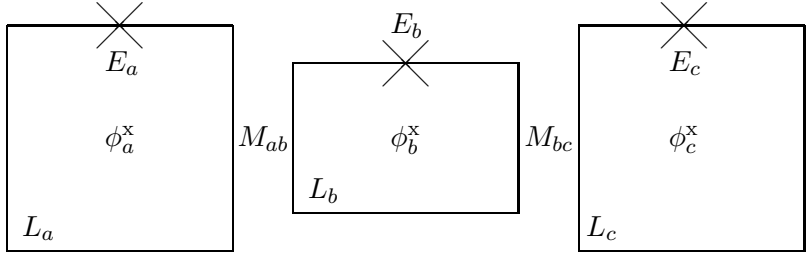

Consider three rf-SQUIDs $a$ through $c$ in a row (Fig.~\ref{fig1}). The $a$- and $c$-devices are supposed to work as (effectively) degenerately biased flux qubits, while the $b$-SQUID is a coupling element, tunable by an external flux bias. To derive the indirect $a$--$c$ coupling it suffices to consider the potential, consisting of both Josephson and magnetic terms ($\hbar=1$ throughout),
\beq
  U=-E_a\cos\phi_a-E_b\cos\phi_b-E_c\cos\phi_c
    +\frac{1}{8e^2}(\bm{\phi}{-}\bm{\phi}^\mathrm{x})^{\!\top}\mathbb{L}^{\!-1}
     (\bm{\phi}{-}\bm{\phi}^\mathrm{x})\;,
\eeql{U}
with the inductance matrix
\beq
  \mathbb{L}=\begin{pmatrix} L_a & -M_{ab} & 0 \\ -M_{ab} & L_b & -M_{bc} \\
                             0 & -M_{bc} & L_c \end{pmatrix}\,.
\eeq
That is, the direct AF inductive $a$--$c$ coupling is ignored. In principle, this can always be achieved through a gradiometric layout. However, as long as $M_{ac}$ is small, one can probably just as well simply add an interaction $M_{ac}I_aI_c$ to the final result (\ref{U-res}) below. In accordance with the roles played by the various loops, the bias fluxes (in phase units) are chosen as
\beq
  \bm{\phi}^\mathrm{x}=\begin{pmatrix} \pi+(M_{ab}/L_b)(\phi_b^{(0)}{-}\phi_b^\mathrm{x}) \\
  \phi_b^\mathrm{x} \\ \pi+(M_{bc}/L_b)(\phi_b^{(0)}{-}\phi_b^\mathrm{x}) \end{pmatrix}\,.
\eeql{phi-x}
Thus, $\phi_{a,c}^\mathrm{x}$ compensate externally for the shielding flux which the $b$-loop couples into the $a$-~and $c$-SQUIDs; see (\ref{phi0b}) for the precise definition of~$\phi_b^{(0)}$.

The $b$-SQUID should act as a passive coupler without introducing additional degrees of freedom, so bistability must be avoided. This can be achieved either by biasing it with a flux close to an integer number of quanta ($\phi_b^\mathrm{x}$ small), or for any flux bias by taking the shielding coefficient $\beta_b\equiv4e^2\!L_bE_b<1$ (in SI units, $\beta=2\pi LI_\mathrm{c}/\Phi_0$). The calculation that follows is valid in either case, and their relative merits will be discussed afterwards.\footnote{A $\beta_b<1$ can clearly be achieved by reducing $E_b$ while keeping the area of the $b$-loop, and hence the inductive coupling, appreciable. If required, a shunting capacitance could be placed across the $b$-junction to keep the $b$-SQUID in the flux regime, where it has the largest response. Also, even a SQUID with a unique potential minimum will have excited states, corresponding to plasma oscillations~$\omega_\mathrm{p}$. While these are outside the present scope, the corresponding excitation energies should exceed any transitions in the $a$- and $c$-devices if the $b$-SQUID is to remain passive. There is an apparent conflict, since the shunting capacitance contemplated above would lower~$\omega_\mathrm{p}$; however, the relevant energy scale, given by the tunnel splittings in qubits $a$ and~$c$, is so small that all requirements can be met simultaneously in practice. Section~\ref{quant} has a few more details on the system's quantum aspects.}

We proceed by expansion in~$M$. Without suggesting that the regime of $M/L\ll1$ (say, distant loops) is the most practical, this leads to a transparent result, which can guide numerical studies in the general case. The $b$-mediated $a$--$c$ coupling is expected to be $\Oc(M^2)$, so the junction phases will be written as
\beq
  \phi_j=\phi_j^{(0)}+\phi^{(1)}_j+\phi^{(2)}_j+\Oc(M^3)\;;
\eeql{phi-exp}
these can be determined by solving the equilibrium condition $\nabla_{\!\!\bm{\phi}}\,U=0$. In leading order, the $\phi_j^{(0)}$ are just the stationary phases of an isolated rf-SQUID,
\begin{align}
  \beta_j\sin\phi_j^{(0)}+\phi_j^{(0)}-\pi&=0\qquad(j=a,c)\;,\\
  \beta_b\sin\phi_b^{(0)}+\phi_b^{(0)}-\phi_b^\mathrm{x}&=0\;,\label{phi0b}
\end{align}
where $\phi_{a,c}^{(0)}-\pi$ can have either sign. To the same order as~(\ref{phi-exp}), one has
\beq
  \mathbb{L}^{\!-1}=\begin{pmatrix} 1/L_a+M_{ab}^2/L_a^2L_b^\pht & M_{ab}/L_aL_b &
    M_{ab}M_{bc}/L_aL_bL_c \\ M_{ab}/L_aL_b &
    1/L_b+M_{ab}^2/L_a^\pht L_b^2+M_{bc}^2/L_b^2L_c^\pht & M_{bc}/L_bL_c \\
    M_{ab}M_{bc}/L_aL_bL_c & M_{bc}/L_bL_c & 1/L_c+M_{bc}^2/L_b^\pht L_c^2\end{pmatrix}
    +\Oc(M^3)\;.
\eeql{Linv}
While (\ref{Linv}) may look tedious, using it consistently actually leads to significant cancellation below.

In $\Oc(M)$, one finds that $\phi^{(1)}_{a,c}=0$ due to the special choice of $\bm{\phi}^\mathrm{x}$ in~(\ref{phi-x}), while the $b$\nobreakdash-loop picks up the shielding fluxes of the neighboring ones, with a prefactor reflecting its susceptibility:
\beq
  \phi^{(1)}_b={\Bigl[1+\beta_b\cos\phi_b^{(0)}\Bigr]}^{-1}
  \biggl[\frac{M_{ab}}{L_a}(\pi{-}\phi_a^{(0)})+\frac{M_{bc}}{L_c}(\pi{-}\phi_c^{(0)})\biggr]\;.
\eeql{chi-b}
Calculating the $\phi^{(2)}_j$ in $\Oc(M^2)$ is straightforward but unnecessary: since one expands around a minimum of~$U$, the $\phi^{(2)}_j$ do not contribute to the relevant order.

All that remains is to substitute (\ref{phi-x})--(\ref{chi-b}) into $U$ in (\ref{U}). Since, e.g., $(\phi_a^{(0)}{-}{\pi)}^2$ does not depend on the qubit state, one is left with
\beq\begin{split}
  U&=\mbox{const}+\frac{M_{ab}M_{bc}}{4e^2L_aL_bL_c}\,
  \frac{\beta_b\cos\phi_b^{(0)}}{1+\beta_b\cos\phi_b^{(0)}}\,
  (\phi_a^{(0)}{-}\pi)(\phi_c^{(0)}{-}\pi)\\
  &=\mbox{const}+M_{ab}M_{bc}\,
  \frac{\beta_b\cos\phi_b^{(0)}}{L_b\bigl(1+\beta_b\cos\phi_b^{(0)}\bigr)}\,I_aI_c\;.
\end{split}\eeql{U-res}
The product of mutual inductances is expected geometrically, while $I_aI_c\propto\sigma_a^z\sigma_c^z$ in qubit parlance. The fraction is worth a closer look:
\beq
  \frac{\beta_b\cos\phi_b^{(0)}}{1+\beta_b\cos\phi_b^{(0)}}\approx\left\{\begin{array}{ll}
  \beta_b\cos\phi_b^\mathrm{x}\;,&\beta_b\ll1\;,\\ 1\;,&\beta_b\gg1\;. \end{array}\right.
\eeql{factor}
Thus, for small $\phi_b^\mathrm{x}$, the coupling is AF, but it changes sign to FM as $\phi_b^\mathrm{x}\To\pi$ (only attainable for $\beta_b<1$).

\begin{figure}[th]
  \includegraphics[width=8cm]{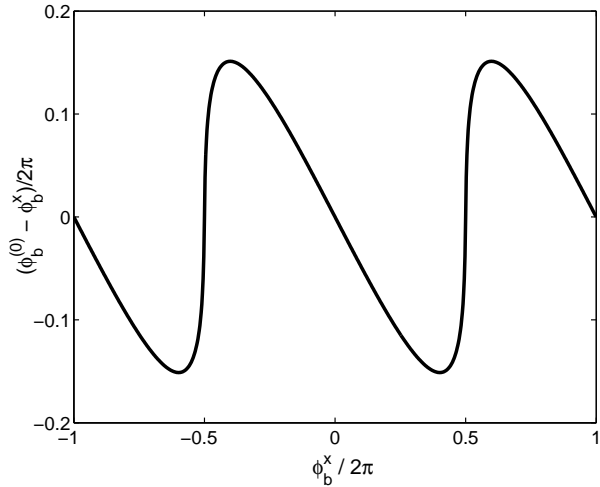}
  \caption{The self-flux $\phi_b^{(0)}(\phi_b^\mathrm{x})-\phi_b^\mathrm{x}$ for an rf-SQUID coupler with $\beta_b=0.95$, according to Eq.~(\ref{phi0b}).}
  \label{icirc}
\end{figure}

This behaviour, and the coupling mechanism in general, have a transparent interpretation. Consider the self-flux curve $\phi_b^{(0)}(\phi_b^\mathrm{x})-\phi_b^\mathrm{x}$ for the free $b$-SQUID (Fig.~\ref{icirc}), which up to constants simply is the shielding current versus applied flux. If the $a$-qubit flips from $\uparrow$ to $\downarrow$, then due to the mutual inductance $-M_{ab}<0$ this effectively increases~$\phi_b^\mathrm{x}$. For small $\phi_b^\mathrm{x}$, the curve has slope $<0$, so that this increase generates a self-flux in the $\downarrow$-direction. In its turn, the latter, through $-M_{bc}<0$, acts as an $\uparrow$-bias for the $c$-loop, favouring the $\uparrow$-state there, i.e., opposite to the state of the $a$-qubit. This also explains the bound of unity exemplified by the second line of (\ref{factor}), since the maximum AF response is perfect shielding for $\beta_b\To\infty$ (an uninterrupted superconducting loop).

However, near $\phi_b^\mathrm{x}=\pi$ the argument works in the opposite direction. There, the self-flux curve has slope $>0$ so that, \emph{differentially}, the self-flux does not shield at all but actually cooperates with the external flux increase. Thus, the coupling changes sign to~FM. Moreover, as $\beta_b\uparrow1$, the slope of this curve increases without bound (a precursor to bistability), so that the FM coupling can actually be large; this corresponds to the zero in the denominator of~(\ref{factor}). On the one hand, the divergence is never realized in practice, since for any finite $M$'s one deals with finite differences, not slopes, on this curve (a nice counterpart to the discussion in~\cite{AB} for charge devices\footnote{Indeed, since the shielding current is the derivative of the energy with respect to the external flux, the induced coupling strength is proportional to the \emph{second} derivative or finite difference of the SQUID band structure, in complete analogy with~\cite{AB}.}). On the other, this potentially large FM coupling makes it possible to overcome any residual direct AF coupling through~$M_{ac}$. Since (\ref{factor}) shows that the large-$\beta_b$ regime is quite inflexible, and that for $\beta_b\ll1$ the coupling strength is limited by the small shielding flux, this case $\beta_b\lesssim1$ is thought to be the preferred one.

The above is mostly a straightforward derivation from (\ref{U}), so very few specifics of the circuit are involved. For instance, the $a$- and $c$-SQUIDs could also be placed inside a large $b$-loop, which changes the signs of both $M_{ab}$ and $M_{bc}$, so that (\ref{U-res}) is invariant. This design modification makes it clearer that the $b$-loop is mostly a flux transformer~\cite{Mooij}, with a weak link providing the tunability. Moreover, the final result (\ref{U-res}) depends only trivially on the properties of the $a$- and $c$-devices---through the flux which these couple into the $b$-SQUID. Hence, the generalization to other types of flux qubits should be obvious.

\subsection{dc-SQUID coupler}
\label{dc}

\begin{figure}[th]
\begin{picture}(100,33)
  \put(0,0){\line(1,0){7}}
  \put(30,0){\line(-1,0){7}}
  \multiput(9,0)(4,0){4}{\oval(4,4)[t]}
  \put(13,4){$L_a$}
  \put(0,0){\line(0,1){30}}
  \put(0,30){\line(1,0){30}}
  \put(30,0){\line(0,1){30}}
  \put(12,27){\line(1,1){6}}
  \put(12,33){\line(1,-1){6}}
  \put(13,14){$\phi_a^\mathrm{x}$}
  \put(13,24){$E_a$}
  \put(35,5){\line(1,0){30}}
  \put(35,5){\line(0,1){2}}
  \multiput(35,9)(0,4){4}{\oval(4,4)[r]}
  \put(38,14){$L_{b1}$}
  \put(35,25){\line(0,-1){2}}
  \put(35,25){\line(1,0){30}}
  \put(65,5){\line(0,1){2}}
  \multiput(65,9)(0,4){4}{\oval(4,4)[l]}
  \put(57,14){$L_{b2}$}
  \put(65,25){\line(0,-1){2}}
  \put(40,22){\line(1,1){6}}
  \put(40,28){\line(1,-1){6}}
  \put(40,29){$E_{b1}$}
  \put(54,22){\line(1,1){6}}
  \put(54,28){\line(1,-1){6}}
  \put(54,29){$E_{b2}$}
  \put(48,14){$\phi_b^\mathrm{x}$}
  \put(50,0){\vector(0,1){3.5}}
  \put(50,0){\line(0,1){5}}
  \put(50,25){\vector(0,1){3.5}}
  \put(50,30){\line(0,-1){5}}
  \put(51,0){$I_b^\mathrm{x}$}
  \put(70,0){\line(1,0){7}}
  \put(100,0){\line(-1,0){7}}
  \multiput(79,0)(4,0){4}{\oval(4,4)[t]}
  \put(82,4){$L_c$}
  \put(70,0){\line(0,1){30}}
  \put(70,30){\line(1,0){30}}
  \put(100,0){\line(0,1){30}}
  \put(81,27){\line(1,1){6}}
  \put(81,33){\line(1,-1){6}}
  \put(82,14){$\phi_c^\mathrm{x}$}
  \put(82,24){$E_c$}
 \end{picture}
\caption{A dc-SQUID mediating a coupling between two rf-SQUID qubits.}
\label{fig2}
\end{figure}
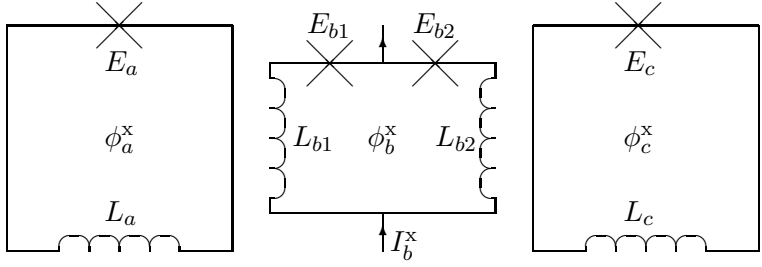

The above indicates that the simple rf-SQUID should be a satisfactory coupling element. Still, it is worthwhile to also examine a dc-SQUID~\cite{Plourde}, which provides additional flexibility since it can be both flux- and current-biased. In particular, if only a current bias~$I_b^\mathrm{x}$ would turn out to suffice for coupling tunability, that could be preferable in situations where a flux bias is problematic due to stray fields etc. The $b$-device now has separate left and right arms, designated as $b1$ and~$b2$, together carrying~$I_b^\mathrm{x}$ (Fig.~\ref{fig2}). Again ignoring the direct $a$--$c$ mutual inductance (not that this makes much difference anywhere), we thus take the coil-to-coil inductance matrix as
\beq
  \begin{pmatrix} L_a      & -M_{ab1} & -M_{ab2} & 0 \\
                  -M_{ab1} & L_{b1}   & M_{b12}  & -M_{b1c} \\
                  -M_{ab2} & M_{b12}  & L_{b2}   & -M_{b2c} \\
                  0        & -M_{b1c} & -M_{b2c} & L_c
  \end{pmatrix}\;.
\eeq
However, always $I_{b2}^\pht=I_{b1}^\pht+I_b^\mathrm{x}$, so the flux--current equation relates three-vectors pertaining to the loops:
\beq
  \bm{\Phi}-\tilde{\bm{\Phi}}^\mathrm{x}=\mathbb{L}\bm{I}\;,
\eeq
with
\begin{gather}
  \bm{\Phi}={(\Phi_a,\Phi_b,\Phi_c)}^{\!\top}\;,\\
  \tilde{\bm{\Phi}}^\mathrm{x}=\bm{\Phi}^\mathrm{x}
    +{(-M_{ab2},\tilde{L}_{b2},-M_{b2c})}^{\!\top} I_b^\mathrm{x}\;,\\
  \bm{I}={(I_a,I_{b1},I_c)}^{\!\top}\;,\\
  \mathbb{L}=\begin{pmatrix} L_a & -M_{ab} & 0 \\ -M_{ab} & L_b & -M_{bc} \\
                             0 & -M_{bc} & L_c \end{pmatrix}\;,\\
  M_{ab}=M_{ab1}+M_{ab2}\;,\qquad M_{bc}=M_{b1c}+M_{b2c}\;,\\
  L_b=\tilde{L}_{b1}+\tilde{L}_{b2}\;,\qquad \tilde{L}_{b1}=L_{b1}+M_{b12}\;,\qquad
    \tilde{L}_{b2}=L_{b2}+M_{b12}\;.
\end{gather}
On the other hand, the four junction phases are all independent dynamical variables. They are related to the fluxes by ${(\phi_a,\phi_{b1}{+}\phi_{b2},\phi_c)}^{\!\top}=2e\bm{\Phi}$.

One should first of all determine the Hamiltonian. Of the Kirchhoff (circuit) equations, the first four are simply the Josephson relations $\dot{\phi}_a=2eQ_a/C_a$ etc., with $Q_a$ the junction charge and $C_a$ its capacitance. The other four express current conservation:
\begin{align}
  \dot{Q}_a&=-2eE_a\sin\phi_a-I_a=-2eE_a\sin\phi_a
    -{[\mathbb{L}^{\!-1}(\bm{\Phi}{-}\tilde{\bm{\Phi}}^\mathrm{x})]}_a\;,\\
  \dot{Q}_{b1}&=-2eE_{b1}\sin\phi_{b1}
    -{[\mathbb{L}^{\!-1}(\bm{\Phi}{-}\tilde{\bm{\Phi}}^\mathrm{x})]}_b\;,\\
  \dot{Q}_{b2}&=-2eE_{b2}\sin\phi_{b2}-I_{b1}^\pht-I_b^\mathrm{x}=-2eE_{b2}\sin\phi_{b2}
    -{[\mathbb{L}^{\!-1}(\bm{\Phi}{-}\tilde{\bm{\Phi}}^\mathrm{x})]}_b-I_b^\mathrm{x}\;,\\
  \dot{Q}_c&=-2eE_c\sin\phi_c
    -{[\mathbb{L}^{\!-1}(\bm{\Phi}{-}\tilde{\bm{\Phi}}^\mathrm{x})]}_c\;.
\end{align}
It is readily seen that these are the Hamilton equations $\dot{\phi}_j=2e\hp\pt H\!/\pt Q_j$ and $\dot{Q}_j=-2e\hp\pt H\!/\pt\phi_j$, for
\beq
  H=\sum_j\biggl[\frac{Q_j^2}{2C_j}-E_j\cos\phi_j\biggr]
  +\half{(\bm{\Phi}{-}\tilde{\bm{\Phi}}^\mathrm{x})}^{\!\top}
  \mathbb{L}^{\!-1}(\bm{\Phi}{-}\tilde{\bm{\Phi}}^\mathrm{x})
  +\frac{I_b^\mathrm{x}}{2e}\phi_{b2}\;,
\eeql{H1}
with $j\in\{a,b1,b2,c\}$. It is less apparent that $H$ is actually symmetric in the two SQUID arms, as it should be: $H$ is a function of the junction phases and charges only, and the asymmetric choice of $I_{b1}$ as the loop current was merely an interim convention. That is, designating the last two terms of (\ref{H1}) as ``magnetic'' and ``bias'' energy, respectively, is a bit arbitrary. Working out~$\mathbb{L}^{\!-1}$, one sees that up to a constant, one can rewrite
\beq
  H=\sum_j\biggl[\frac{Q_j^2}{2C_j}-E_j\cos\phi_j\biggr]
    +\half{(\bm{\Phi}{-}\bm{\Phi}^\mathrm{x})}^{\!\top}\mathbb{L}^{\!-1}
    (\bm{\Phi}{-}\bm{\Phi}^\mathrm{x})+H_\mathrm{bias}\;,
\eeql{H2}
where
\beq\begin{split}
  \smash{\frac{2e\det{\mathbb{L}}}{I_b^\mathrm{x}}}H_\mathrm{bias}&=
  [(M_{ab2}\tilde{L}_{b1}{-}M_{ab1}\tilde{L}_{b2})L_c
    +(M_{ab1}M_{b2c}{-}M_{ab2}M_{b1c})M_{bc}]\phi_a\\
  &\quad+[M_{ab2}M_{ab}L_c+L_aM_{b2c}M_{bc}-L_a\tilde{L}_{b2}L_c]\phi_{b1}\\
  &\quad+[-M_{ab1}M_{ab}L_c-L_aM_{b1c}M_{bc}+L_a\tilde{L}_{b1}L_c]\phi_{b2}\\
  &\quad+[L_a(\tilde{L}_{b1}M_{b2c}{-}\tilde{L}_{b2}M_{b1c})
    +M_{ab}(M_{ab2}M_{b1c}{-}M_{ab1}M_{b2c})]\phi_c\;,
\end{split}\eeql{Hb}
with $\det{\mathbb{L}}=L_aL_bL_c-L_aM_{bc}^2-M_{ab}^2L_c$. It would be instructive to see if the above, especially the form (\ref{Hb}) for $H_\mathrm{bias}$, can be reproduced (presumably almost mechanically) by the ``node formalism'' of circuit analysis~\cite{devoret,burkard}.

From here on, the analysis is a straightforward generalization of the rf-SQUID case, again writing the phases as in (\ref{phi-exp}), and expanding $H_\mathrm{bias}$ in (\ref{Hb}) to the same order. We again want the $a$- and $c$-qubits to be effectively at degeneracy, which presently means that their own external flux should compensate also for~$I_b^\mathrm{x}$:
\beq
  \bm{\phi}^\mathrm{x}=\begin{pmatrix}
    \pi+L_b^{-1}[M_{ab}(\phi_{b1}^{(0)}{+}\phi_{b2}^{(0)}{-}\phi_b^\mathrm{x})
    +2eI_b^\mathrm{x}(M_{ab2}\tilde{L}_{b1}{-}M_{ab1}\tilde{L}_{b2})] \\
  \phi_b^\mathrm{x} \\
  \pi+L_b^{-1}[M_{bc}(\phi_{b1}^{(0)}{+}\phi_{b2}^{(0)}{-}\phi_b^\mathrm{x})
    +2eI_b^\mathrm{x}(\tilde{L}_{b1}M_{b2c}{-}\tilde{L}_{b2}M_{b1c})] \end{pmatrix}\,.
\eeql{phi-x-dc}

In $\Oc(M^0)$, the stationary phases obey the standard equations for the isolated devices:
\begin{align}
  \beta_j\sin\phi_j^{(0)}+\phi_j^{(0)}-\pi&=0\qquad(j=a,c)\;,\\
  \beta_{b1}\sin\phi_{b1}^{(0)}+\phi_{b1}^{(0)}+\phi_{b2}^{(0)}-\phi_b^\mathrm{x}
    -2eI_b^\mathrm{x}\tilde{L}_{b2}^\pht&=0\;,\label{phi0b1}\\
  \beta_{b2}\sin\phi_{b2}^{(0)}+\phi_{b1}^{(0)}+\phi_{b2}^{(0)}-\phi_b^\mathrm{x}
    +2eI_b^\mathrm{x}\tilde{L}_{b1}^\pht&=0\;.\label{phi0b2}
\end{align}
Here, $\beta_{a(c)}\equiv4e^2L_{a(c)}E_{a(c)}$ is standard, but note that the definitions $\beta_{bj}\equiv4e^2L_{b}E_{bj}$ ($j=1,2$) involve the full loop inductance of the $b$-SQUID, not the individual arm inductances.

In $\Oc(M)$, our choice (\ref{phi-x-dc}) for $\bm{\phi}^\mathrm{x}$ again ensures that $\phi^{(1)}_a=\phi^{(1)}_c=0$, while the equations for $\phi^{(1)}_{b1(2)}$ are coupled, as were (\ref{phi0b1})--(\ref{phi0b2}) for $\phi_{b1(2)}^{(0)}$ (although, unlike the latter, they are of course linear):
\beq
  \begin{pmatrix} \phi^{(1)}_{b1} \\ \phi^{(1)}_{b2} \end{pmatrix} =
  {\begin{pmatrix} 1{+}\beta_{b1}\cos\phi_{b1}^{(0)} & 1 \\ 1 & 1{+}\beta_{b2}\cos\phi_{b2}^{(0)}
  \end{pmatrix}}^{\!\!-1}
  \biggl[\frac{M_{ab}}{L_a}(\pi{-}\phi_a^{(0)})+\frac{M_{bc}}{L_c}(\pi{-}\phi_c^{(0)})\biggr]
  \begin{pmatrix} 1 \\ 1 \end{pmatrix}\,.
\eeql{chi-dc}

Like for the rf-SQUID, it is best to leave the $\phi^{(2)}_j$ unevaluated in the $M$-expansion of~$U$, and one sees that their contribution cancels (to this order), since one expands around a potential minimum. Taking advantage of some further cancellation and dropping all terms which do not depend on the qubit state, one arrives at
\beq
  U=\mbox{const}+M_{ab}M_{bc}\,
  \frac{\beta_{b1}\beta_{b2}\cos\phi_{b1}^{(0)}\cos\phi_{b2}^{(0)}}
    {L_b(\beta_{b1}\cos\phi_{b1}^{(0)}+\beta_{b2}\cos\phi_{b2}^{(0)}
      +\beta_{b1}\beta_{b2}\cos\phi_{b1}^{(0)}\cos\phi_{b2}^{(0)})}
  \,I_aI_c\;,
\eeql{U-res-dc}
in which all tunability thus comes via the dependence of $\phi_{b1(2)}^{(0)}$ on $(\phi_b^\mathrm{x},I_b^\mathrm{x})$ as given in \mbox{(\ref{phi0b1})--(\ref{phi0b2})}. For a simple consistency check, let us take $I_b^\mathrm{x}\To0$ and $E_{b2}\To\infty$. Then, (\ref{phi0b2}) shows that $\phi_{b2}^{(0)}=0$, while (\ref{phi0b1}) reduces to $\beta_{b1}\sin\phi_{b1}^{(0)}+\phi_{b1}^{(0)}-\phi_b^\mathrm{x}=0$, and the large fraction in (\ref{U-res-dc}) becomes $\beta_{b1}\cos\phi_{b1}^{(0)}/\bigl[1+\beta_{b1}\cos\phi_{b1}^{(0)}\bigr]$; in view of the remark below (\ref{phi0b2}), this is exactly the result for rf-SQUID coupling.

Another well-known case is the symmetric dc-SQUID with negligible shielding, for which (\ref{phi0b1})--(\ref{phi0b2}) reduce to $\phi_{b1}^{(0)}+\phi_{b2}^{(0)}=\phi_b^\mathrm{x}$ and $2I_\mathrm{c}\cos(\phi_b^\mathrm{x}/2)\sin(\phi_{b1}^{(0)}{-}\phi_b^\mathrm{x}/2)= I_b^\mathrm{x}$. Even in this simple case one can have FM coupling, for $\cos^2(\phi_b^\mathrm{x}/2)<|I_b^\mathrm{x}|/2I_\mathrm{c}< \left|\cos(\phi_b^\mathrm{x}/2)\right|$. While one thus needs a nonzero flux bias, tunability of $I_b^\mathrm{x}$ suffices for changing the sign of the coupling.

One notes a few features, and it should be investigated whether these carry over to more generic devices described by our equations: (a)~to achieve FM coupling, one needs nonzero flux \emph{and} current bias; (b)~the denominator in (\ref{U-res-dc}) is always positive; (c)~$\cos\phi_{b1}^{(0)}$ and $\cos\phi_{b2}^{(0)}$ never become negative simultaneously. \emph{Need for current bias}: for a symmetric SQUID this is easy to prove, since for $I_b^\mathrm{x}=0$ one then has $\phi_{b1}^{(0)}=\phi_{b2}^{(0)}$, and for any $\phi_b^\mathrm{x}$ there is a stationary state with $|\phi_{b1}^{(0)}|\le\pi\hn/2$. This can\emph{not} be generalized further, since we have already seen that the (asymmetric) dc-SQUID contains the rf-SQUID as a special case, and the latter can mediate FM coupling without any current bias. \emph{Need for flux bias}: again easily proved for symmetric devices with arbitrary shielding, which have $\phi_{b1}^{(0)}=-\phi_{b2}^{(0)}$ etc.\ for $\phi_b^\mathrm{x}=0$; in general, however, (\ref{phi0b1})--(\ref{phi0b2}) show that an inductance imbalance between the two SQUID arms can play the same role as a nonzero~$\phi_b^\mathrm{x}$. Finally, note that the rhs in (\ref{chi-dc}) features the second-derivative matrix for the potential of the free $b$-SQUID (characterizing the device's susceptibility). For a stable minimum, this matrix should be positive-definite, from which (b) and (c) immediately follow in general. (As remarked above, the second derivatives keep appearing because of the $M$-expansion. For finite~$M$, one has second differences instead~\cite{AB}.)

One of the possible uses for a dc-SQUID coupler could be to moonlight as a readout device through switching-current measurement, so let us treat that in the present notation. Rather than think of the maximum bias for which (\ref{phi0b1})--(\ref{phi0b2}) have solutions, it is convenient to fix $\phi_b^\mathrm{x}$ and consider $I_b^\mathrm{x}$ as a function of the junction phases, which are still related among themselves by one nontrivial constraint. Critical bias now corresponds to $dI_b^\mathrm{x}/d\phi=0$; working things out readily yields
\beq
  \beta_{b1}\cos\phi_{b1}^{(0)}+\beta_{b2}\cos\phi_{b2}^{(0)}
      +\beta_{b1}\beta_{b2}\cos\phi_{b1}^{(0)}\cos\phi_{b2}^{(0)}=0
\eeql{Ib-crit}
as the relevant condition.\footnote{Compare (\ref{U-res-dc}) and the end of the previous paragraph: one can equivalently characterize critical bias as the current for which the potential becomes marginally unstable. We note in passing that, for near-critical bias, the induced coupling is always FM, except for the doubly nongeneric points given by (\ref{Imax}) and~(\ref{Imin}).} To calculate switching-current sensitivity, we now vary~$\phi_b^\mathrm{x}$, and see which variation in $I_b^\mathrm{x}$ conserves the above relation. One finds
\beq
  \frac{dI_b^\mathrm{x,max}}{d\phi_b^\mathrm{x}}=
  \frac{\beta_{b1}\cos\phi_{b1}^{(0)}}
       {2e\bigl[L_b+\tilde{L}_{b1}\beta_{b1}\cos\phi_{b1}^{(0)}\bigr]}\;;
\eeql{sens}
using (\ref{Ib-crit}), one again finds that this is more symmetric in the SQUID arms than it looks. So the sensitivity vanishes iff, at critical bias, both junctions become critical individually. This happens for
\beq
  \phi_b^\mathrm{x}=\pm\frac{\tilde{L}_{b1}\beta_{b1}-\tilde{L}_{b2}\beta_{b2}}{L_b}\;,
  \qquad I_b^\mathrm{x}=\pm\frac{\beta_{b1}+\beta_{b2}}{2eL_b}\;.
\eeql{Imax}
As long as the ``critical flux" product $\tilde{L}_{bj}I_{\mathrm{c},bj}$ is different for the two arms, one has a finite sensitivity at $\phi_b^\mathrm{x}=0$ and in fact for any $\phi_b^\mathrm{x}$, since one can ramp $I_b^\mathrm{x}$ with either sign. For the sake of completeness we also give the zero-sensitivity points
\beq
  \phi_b^\mathrm{x}=
    \pm\biggl[\pi-\frac{\tilde{L}_{b1}\beta_{b1}+\tilde{L}_{b2}\beta_{b2}}{L_b}\biggr]\;,
  \qquad I_b^\mathrm{x}=\pm\frac{\beta_{b2}-\beta_{b1}}{2eL_b}\;,
\eeql{Imin}
which correspond to \emph{minima} in the switching current.

Finally, the above analysis only shows that the potential part of (\ref{H2})--(\ref{Hb}) enables tunable coupling in the classical case. That is, there could be devices outside its regime of validity which are nonetheless suitable couplers, as long as the transformer remains passive, i.e., confined to its lowest energy state/band~\cite{AB}. For a quantum analysis, the interacting, biased Hamiltonian (\ref{H2})--(\ref{Hb}) can be used as a starting point. The method of $M$\nobreakdash-expansion should again reduce the problem to one for the uncoupled, biased, dc-SQUID. Below, however, the quantum case is taken up for rf-SQUIDs only.

\section{Quantum analysis}
\label{quant}

Given that quantum tunneling is crucial for the qubits $a$ and~$c$, a consistent quantum treatment of the whole system is desirable. In the quantum case, the $M$-expansion works as above. The other crucial step is an adiabatic approximation for the $b$-coupler (cf.~\cite{AB}), replacing operators with their ground-state expectation. This very language is only meaningful if the rf-SQUID $b$ is a well-defined separate entity, i.e., for weak coupling. Specifically, the coupling energies must be much smaller than the first excitation energy of the $b$-device, but this does not restrict their size compared to the level splittings of qubits $a$ and~$c$.

Thus, let us collect all terms in the Hamiltonian $H=\sum_{j=a,b,c}Q_j^2/2C_j-U$ [with $U$ as in~(\ref{U})] involving the coupler. To the relevant order in~$M$ one has
\begin{align}
  H_b&\equiv\frac{Q_b^2}{2C_b}-E_b\cos\phi_b+\frac{1}{8e^2L_b}\biggl\{
  \left(1+\frac{M_{ab}^2}{L_aL_b}+\frac{M_{bc}^2}{L_bL_c}\right)
  (\phi_b^\pht{-}\phi_b^\mathrm{x})^2\notag\\
  &\quad\,+2\frac{M_{ab}}{L_a}(\phi_a^\pht{-}\phi_a^\mathrm{x})
  (\phi_b^\pht{-}\phi_b^\mathrm{x})+2\frac{M_{bc}}{L_c}(\phi_b^\pht{-}\phi_b^\mathrm{x})
  (\phi_c^\pht{-}\phi_c^\mathrm{x})\biggr\}\\
  &=\frac{Q_b^2}{2C_b}-E_b\cos\phi_b
  +\frac{(\phi_b^\pht{-}\phi_b^\mathrm{x,eff})^2-(\delta\hn\phi_b^\mathrm{x})^2}
  {8e^2L_b^\mathrm{eff}}\;,
  \intertext{with}
  L_b^\mathrm{eff}&=
  L_b\left(1-\frac{M_{ab}^2}{L_aL_b}-\frac{M_{bc}^2}{L_bL_c}\right)\;,\\
  \phi_b^\mathrm{x,eff}&=\phi_b^\mathrm{x}+\delta\hn\phi_b^\mathrm{x}
  =\phi_b^\mathrm{x}+\frac{M_{ab}}{L_a}(\phi_a^\mathrm{x}{-}\phi_a^\pht)
  +\frac{M_{bc}}{L_c}(\phi_c^\mathrm{x}{-}\phi_c^\pht)\;.
\end{align}
The adiabatic step amounts to the replacement
\beq
  H_b\mapsto\Ec_b^{(0)}(\phi_b^\mathrm{x,eff})
  -\frac{(\delta\hn\phi_b^\mathrm{x})^2}{8e^2L_b}
\eeql{Hb-adia}
and for small coupling, one can further write
\beq
  \Ec_b^{(0)}(\phi_b^\mathrm{x,eff})\approx\Ec_b^{(0)}(\phi_b^\mathrm{x})
  -\frac{I_b^{(0)}}{2e}\delta\hn\phi_b^\mathrm{x}
  +\frac{\chi_b^{(0)}}{8e^2}(\delta\hn\phi_b^\mathrm{x})^2\;.
\eeql{adia2}
Here, $I_b^{(0)}$ is the ground-state expectation of the loop current; the susceptibility is defined as
\beq
  \chi_b^{(0)}=-\frac{dI_b^{(0)}}{d\Phi_b^\mathrm{x}}
  =\frac{1}{L_b}\biggl(1-
    \frac{d{\langle\phi_b\rangle}_0}{d\phi_b^\mathrm{x}}\biggr)\;,
\eeql{chi}
which gives the bound $\chi_b^{(0)}<1/L_b$ (no absolute value on the lhs)---the maximum shielding is perfect diamagnetism for an uninterrupted loop [cf.\ below~(\ref{factor})].

All that remains is to substitute (\ref{Hb-adia}) and (\ref{adia2}) back into the full $H$; after some cancellation, one finds
\begin{align}
  H&=\frac{Q_a^2}{2C_a}-E_a\cos\phi_a
  +\frac{(\phi_a{-}\phi_a^\mathrm{x,eff})^2}{8e^2L_a^\mathrm{eff}}
  +(a\leftrightarrow c)+H_\mathrm{int}\;,\\
  \phi_a^\mathrm{x,eff}&=\phi_a^\mathrm{x}-2eI_b^{(0)}M_{ab}\;,\label{phix-eff}\\
  L_a^\mathrm{eff}&=L_a\left(1-\frac{M_{ab}^2}{L_a}\chi_b^{(0)}\right)\;,\label{La-eff}\\
  H_\mathrm{int}&=M_{ab}M_{bc}\chi_b^{(0)}I_aI_c\;.\label{Hint}
\end{align}
This does not involve assumptions on the states of the $a$- and $c$-qubits, or on their flux biases. Note that (\ref{Hint}) involves current \emph{operators} for the $a$- and $c$-devices~\cite{H2qb}, in contrast to, e.g., (\ref{chi}) and (\ref{phix-eff}). Equations (\ref{phix-eff}) and~(\ref{La-eff}) imply that the qubits' effective parameters depend subtly on the coupler's working point, presenting a possible experimental challenge.

As in the classical case, the perturbative step has reduced the problem to that of the uncoupled $b$-SQUID, and we henceforth omit its index as well as the zero superscript indicating free devices. Evaluating (\ref{chi}) in the potential minimum, one verifies that (\ref{Hint}) reduces to (\ref{U-res}) in the classical limit. In the quantum case, numerical differentiation in (\ref{chi}) presents no problems, but it will be instructive, and useful for Sec.~\ref{bist}, to express $\chi$ in terms of quantities at a single bias point. To this end, (\ref{chi}) is evaluated in terms of $\xi_0(\phi)\equiv\partial_{\phi^\mathrm{x}}\psi_0(\phi)$, where normalization implies that $\langle\psi_0|\xi_0\rangle=0$. In its turn, $\xi_0$ is solved from the $\phi^\mathrm{x}$-differentiated Schr\"odinger equation
\beq
  (H-\Ec_0)\xi_0+\biggl(
  \frac{\phi^\mathrm{x}{-}\phi}{4e^2L}-\Ec_0'\biggr)\psi_0=0\;,
\eeql{diff-Sch}
in which $\psi_0$ figures as an inhomogeneous term, and where $\Ec_0'\equiv\partial_{\phi^\mathrm{x}}\Ec_0= {\langle\phi^\mathrm{x}{-}\phi\rangle}_0/4e^2L$.
Using a variation-of-constant solution to (\ref{diff-Sch}), one then derives\footnote{All wave functions are taken real w.l.o.g.}
\beq
  \chi=\frac{1}{L}\biggl[1-\frac{C}{4e^4L}\int_{-\infty}^\infty
    \frac{d\phi}{\psi_0^2(\phi)}\biggl\{\int_{-\infty}^\phi\!\!d\phi_1
    \bigl(\phi_1{-}{\langle\phi_1\rangle}_0\bigr)\psi_0^2(\phi_1)
    \biggr\}^{\!\!2}\biggr]\;.
\eeql{chi-res1}
This already expresses considerable simplification, in that two of the integrals in a triple integral were written as the square of $k(\phi)\equiv\int_{-\infty}^\phi d\phi_1\, \bigl(\phi_1{-}{\langle\phi_1\rangle}_0\bigr)\psi_0^2(\phi_1)=\linebreak -\int_\phi^\infty d\phi_1\, \bigl(\phi_1{-}{\langle\phi_1\rangle}_0\bigr)\psi_0^2(\phi_1)$. However, $k(\phi)\To0$ exponentially for $\phi\To\infty$ ($\phi\To-\infty$) will not be realized numerically for the former (latter) representation of~$k$. Hence, we have to use both. Putting $l_1(\zeta)\equiv\int_{-\infty}^\zeta d\phi\,k^2(\phi)/\psi_0^2(\phi)$ and $l_2(\zeta)\equiv\int_\zeta^\infty d\phi\,k^2(\phi)/\psi_0^2(\phi)$, one has $\chi=L^{-1}[1-(C/4e^4L)\{l_1(0)+l_2(0)\}]$. Now, $\chi$ can be found by a \emph{single} integration of the ODE systems
\beq
  \left\{\!\!\begin{array}{l} l_1'=k^2/\psi_0^2 \\
  k'(\phi)=(\phi-{\langle\phi\rangle}_0\bigr)\psi_0^2(\phi) \\
  l_1(-\infty)=k(-\infty)=0 \end{array}\right.\;,\qquad
  \left\{\!\!\begin{array}{l} l_2'=-k^2/\psi_0^2 \\
  k'(\phi)=(\phi-{\langle\phi\rangle}_0\bigr)\psi_0^2(\phi) \\
  l_2(\infty)=k(\infty)=0 \end{array}\right.\;.\label{ODE2}
\eeq

A different derivation uses that, for \emph{stationary} states, the shielding current equals the Josephson current (readily proven formally):
\beq
  \chi=4e^2E\frac{d\langle\sin\phi\rangle_0}{d\phi^\mathrm{x}}\;.
\eeql{chi2}
It is correct to evaluate (\ref{chi2}) with the aid of (\ref{diff-Sch}), but this yields unwieldy asymmetric expressions. Instead, we observe that a change in flux bias is a uniform translation if there is no Josephson potential, i.e., we set $\xi_0\equiv\eta_0-\psi_0'$ with $\langle\psi_0|\eta_0\rangle=0$ and
\beq
  (H-\Ec_0)\eta_0+(E\sin\phi-\Ec_0')\psi_0=0\;,
\eeq
using the matching expression $\Ec_0'=E{\langle\sin\phi\rangle}_0$. Combination leads to
\beq
  \chi=4e^2E{\langle\cos\phi\rangle}_0-4CE^2\int_{-\infty}^\infty
    \frac{d\phi}{\psi_0^2(\phi)}\biggl\{\int_{-\infty}^\phi\!\!d\phi_1
    \bigl(\sin\phi_1{-}{\langle\sin\phi_1\rangle}_0\bigr)\psi_0^2(\phi_1)
    \biggr\}^{\!\!2}\;.
\eeql{chi-res2}
The translation of (\ref{chi-res2}) to ODE form is wholly analogous to (\ref{ODE2}) and will not be repeated.

For small~$\beta$, (\ref{chi-res2}) is clearly preferable to (\ref{chi-res1}), since the double integral yields only a small correction to the first term. Compare (\ref{chi2}), readily yielding an estimate of $\chi$ for weakly shielded SQUIDs, to (\ref{chi}), where in this case one has to evaluate the difference of two nearly equal terms. Conversely, for large $\beta$, ${\langle\phi\rangle}_0\approx2\pi n$ is approximately constant over most of the bias range, giving a clear advantage to (\ref{chi}) and (\ref{chi-res1}), which capture this behaviour in their first terms. The equivalence of the four expressions for $\chi$ has been thoroughly tested numerically.

\begin{figure}
  \includegraphics[width=5cm]{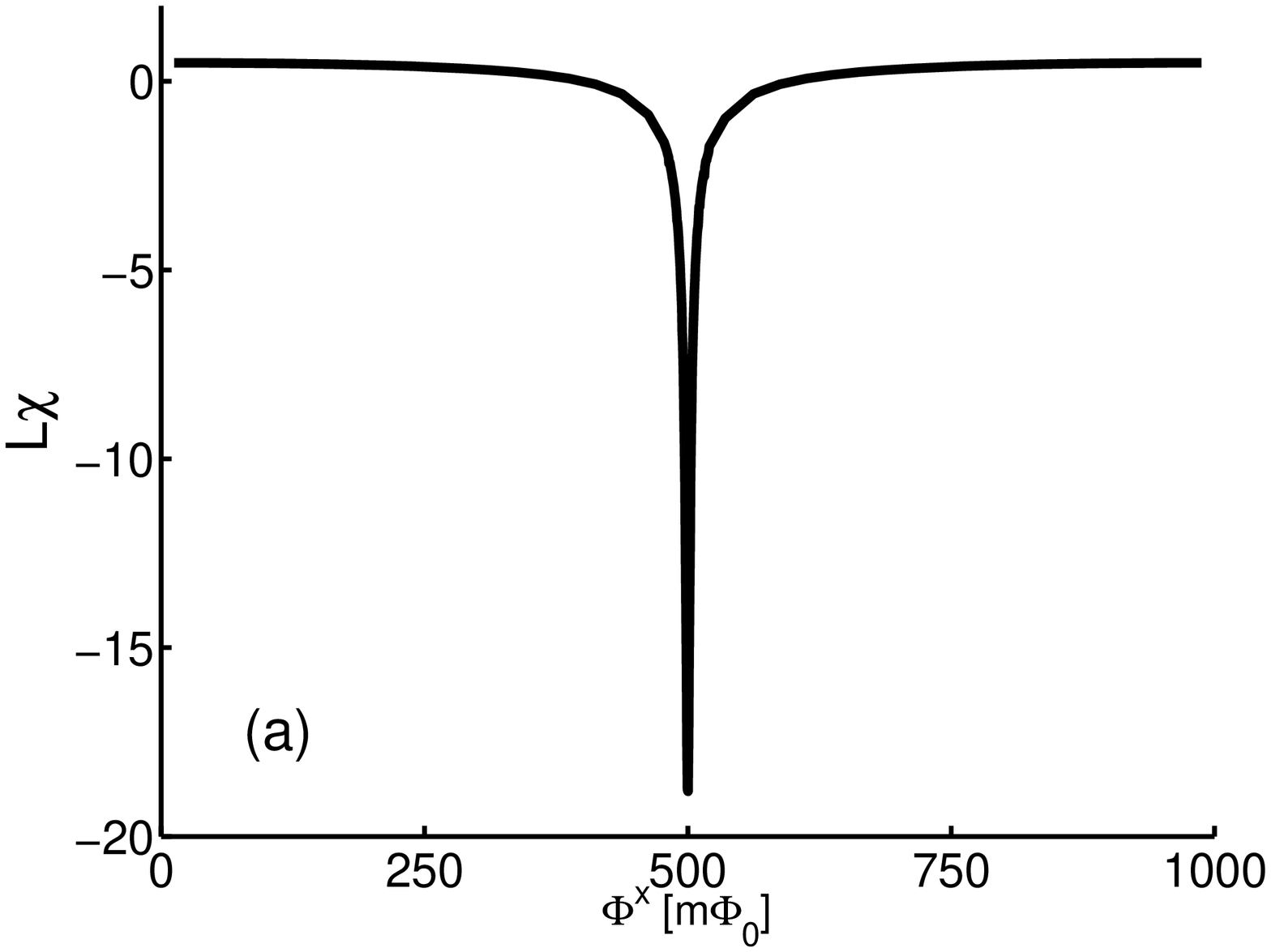}
  \includegraphics[width=5cm]{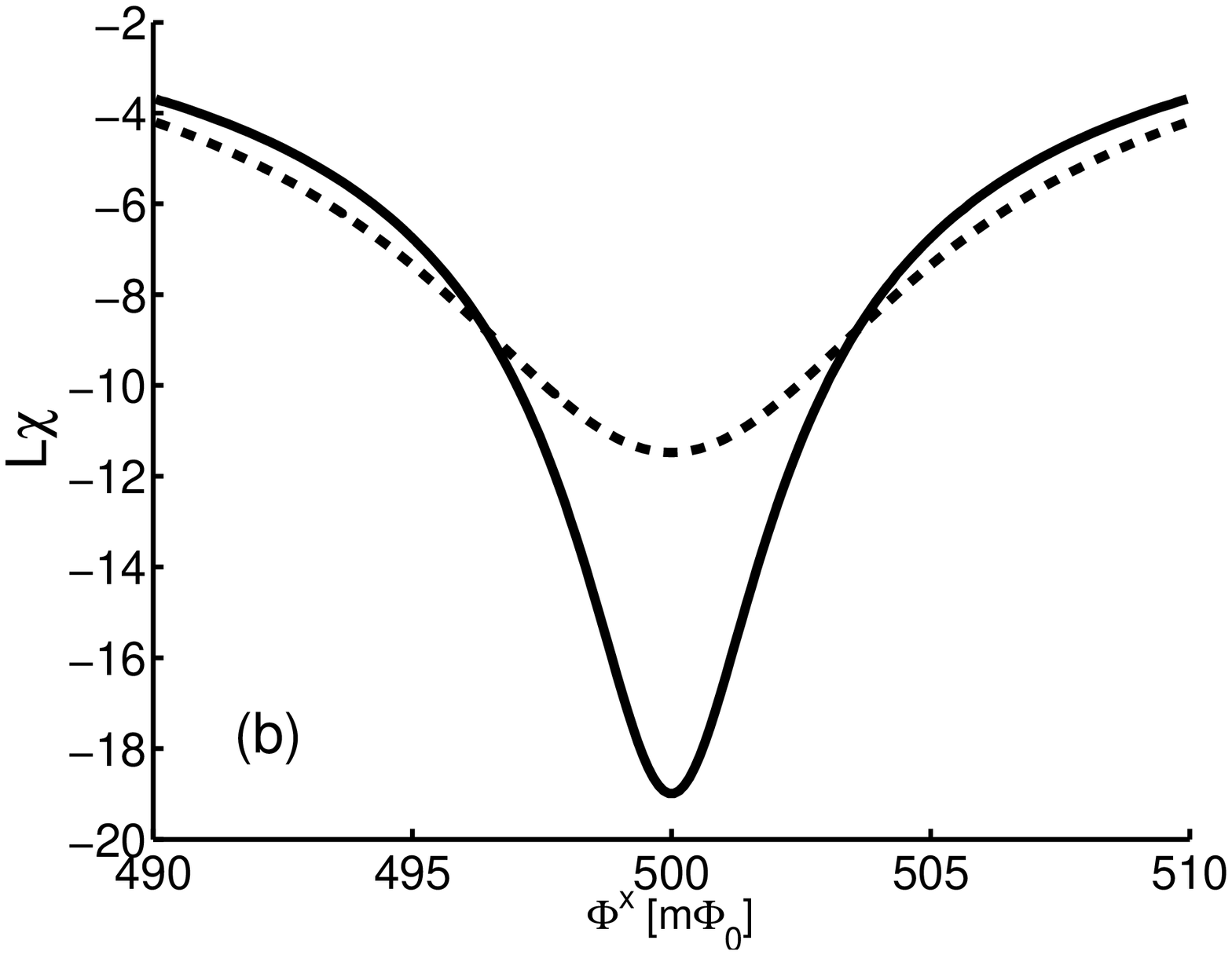}
  \includegraphics[width=5cm]{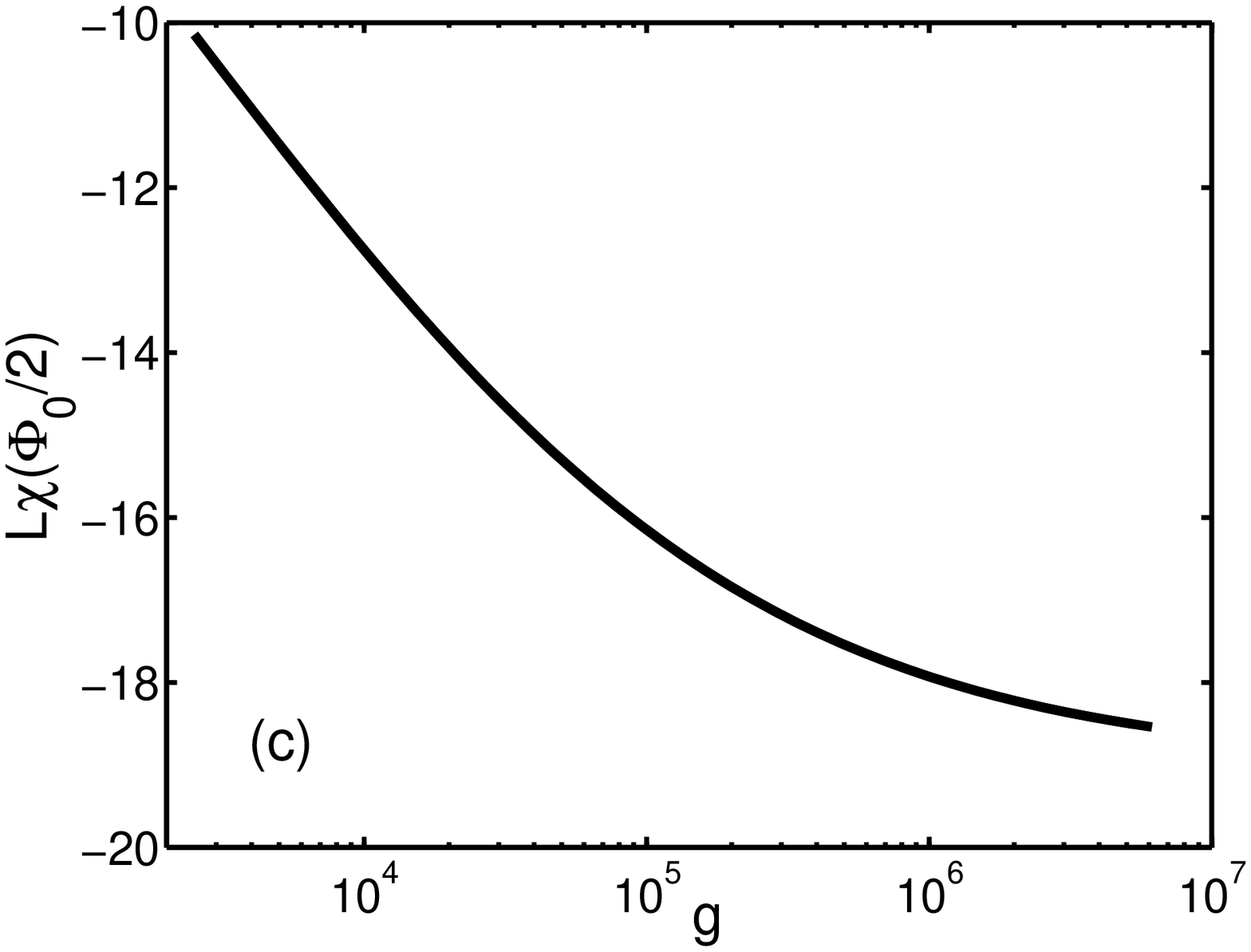}
  \caption{The reduced SQUID susceptibility $L\chi$ versus flux bias $\Phi^\mathrm{x}$ for~$\beta=0.95$. (a)~Classical response. (b)~Close-up of the classical (solid line) and quantum response (dashed line, for $g=5\cdot10^3$) for biases near $\frac{1}{2}\Phi_0$; the two curves are indistinguishable elsewhere. (c)~The quantum susceptibility at $\Phi^\mathrm{x}=\frac{1}{2}\Phi_0$ versus capacitance, showing the approach to the classical limit.}
  \label{chi-plots}
\end{figure}

Results are shown in Fig.~\ref{chi-plots}. It is seen that the finite spread of the ground-state wave function smooths out the classical susceptibility, in particular near $\phi^\mathrm{x}=\pi$ where the FM response is maximal. The finite quantum tunneling amplitude also means that there is no longer a sharp transition at $\beta=1$, although the small tunnel splitting for larger $\beta$ is likely to violate the adiabaticity requirement.

\section{Application: SQUID susceptibility near an anticrossing}
\label{bist}

\subsection{Analytics}
\label{bist-a}

While the above was derived with monostable coupling devices in mind, it should be general, hence equally applicable to rf-SQUID qubits close to the anticrossing, where it can be compared to the predictions of the two-state model
\beq
  \Ec_0^\mathrm{qb}(\delta\Phi^\mathrm{x})=\text{const} -
  \sqrt{\Delta(\delta\Phi^\mathrm{x})^2+\epsilon(\delta\Phi^\mathrm{x})^2}\;,
\eeql{Eqb}
with $\delta\Phi^\mathrm{x}\equiv\Phi^\mathrm{x}-\frac{1}{2}\Phi_0$. The bias $\epsilon$ is defined as half the difference between the ``local well energies", which for definiteness are taken as the ground-state energies with an artificial nodal condition imposed at the potential maximum. If $\beta-1$ is taken fixed, then the two-state model should be valid asymptotically in $g\equiv 2CE/e^2$. Note that it incorporates several quantum effects short of tunneling: at finite bias the curvatures of the two wells change in opposite directions, affecting the zero-point energies (note~15 in Ref.~\cite{IMT}), and the anharmonicity of the wells is also accounted for. The ``tunneling amplitude'' $\Delta$ is primarily defined through (\ref{Eqb}), without reliance on WKB estimates. Bearing in mind that $\Delta(\delta\Phi^\mathrm{x})$ [$\epsilon(\delta\Phi^\mathrm{x})$] is even [odd], one finds $\chi^\mathrm{qb}(0)=\Ec_0^\mathrm{qb}(0)''= -\Delta''(0)-I_\mathrm{p}^2/\Delta$ for the two-state approximation to~$\chi$,
with $\Delta\equiv\Delta(0)$ and $I_\mathrm{p}\equiv-\epsilon'(0)$.

These predictions make it clear that $\chi(0)$ is exponentially large in~$g$, so that in the evaluation of (\ref{chi-res1}) and (\ref{chi-res2}), their first terms can be safely ignored. Further, ${\langle\phi\rangle}_0=\pi$ and ${\langle\sin\phi\rangle}_0=0$ by symmetry for $\phi^\mathrm{x}=\pi$, which will be assumed henceforth. In either formula, the increase in $\chi$ can only come from the factor $\psi_0^{-2}$, which is large for $|\phi|\To\infty$ and near the barrier. However, in both formulas and both for $\phi\To\infty$ and $\phi\To-\infty$, this is overcome by $\{\cdots\}^2$ decaying as $\psi_0^4$. Therefore, the only relevant contribution comes from the barrier region, where to within exponentially small corrections the factors $\{\cdots\}^2$ are constant, and readily related to the loop current. Thus, in the regime of interest, both (\ref{chi-res1}) and (\ref{chi-res2}) evaluate to
\beq
  \chi(0)\approx
  -\frac{CI_\mathrm{p}^2}{4e^2}\int_\mathrm{b}\frac{d\phi}{\psi_0^2(\phi)}\;,
\eeql{chi0-int}
where $\int_\mathrm{b}$ means an integral over the barrier region, which for definiteness can be taken as everything between the two well minima (so as not to rely on ``classical turning points''). To relate the integral in (\ref{chi0-int}) to~$\Delta$, define $f(\phi,\Ec)$ as the solution to the Schr\"odinger equation at (continuously tunable) energy $\Ec$ satisfying the left boundary condition. Of course, this function will diverge for $\phi\To\infty$ unless $\Ec$ is an eigenenergy. Its energy derivative satisfies a Schr\"odinger equation in which $f$ figures as an inhomogeneous term [cf.\ (\ref{diff-Sch}) for the bias derivative], which we solve by the variation-of-constant Ansatz $\partial_\Ec f=fh$. Assuming that the $\Ec$-dependence of $f$ can be linearized for $\Ec_0\le\Ec\le\Ec_1$, the significance of these functions is that $\partial_\Ec f(\pi,\Ec)\approx[0-\psi_0(\pi)]/2\Delta\;\Rightarrow\; h(\pi,\Ec_0)\approx-1/2\Delta$. Solving for $h$, one finds
\beq
  h'(\phi)=-\frac{C}{2e^2}\int_{-\infty}^\phi\!\!\!d\phi_1\,
  \frac{f^2(\phi_1)}{f^2(\phi)}\;;
\eeq
this defines $h$ only up to a constant, since $f$ was defined only up to a, possibly $\Ec$-dependent, normalization. We now fix $\int_{-\infty}^\pi\!d\phi\,f^2(\phi)=\frac{1}{2}$, so that $h'(\phi)\approx-C/4e^2f^2(\phi)$ throughout the barrier region. Further, with $\phi_\mathrm{min}$ the left potential minimum, $h(\phi_\mathrm{min},\Ec)$ is of order one, hence negligible, and one finds $h(\pi,\Ec_0)\approx-(C/4e^2)\int_{\phi_\mathrm{min}}^\pi\!d\phi\,\psi_0^{-2}(\phi) =-(C/8e^2)\int_\mathrm{b}\!d\phi\,\psi_0^{-2}(\phi)$.

Combination shows that $\Delta^{-1}=(C/4e^2)\int_\mathrm{b}\!d\phi\,\psi_0^{-2}(\phi)$, and substitution in (\ref{chi0-int}) yields
\beq
  \chi(0)\approx-\frac{I_\mathrm{p}^2}{\Delta}\;.
\eeq
Comparing with what was found below (\ref{Eqb}) for $\chi^\mathrm{qb}(0)$ in the two-state approximation, it leads one to conclude that actually $\Delta''(0)=0$. More precisely, comparing the orders of magnitude, it indicates that $\Delta$ does not have relative variations of order one in the interval $|\epsilon|\sim\Delta$, which is all one cares about. This flatness of $\Delta$ in the anticrossing region is often assumed without much discussion, let alone proof; even the comparatively detailed WKB treatment in Ref.~\cite{AFL} is complete only for highly excited states in the local wells. Yet, the standard WKB expression for the tunneling amplitude is very sensitive to the barrier action integral, suggesting that there could be an equally strong sensitivity to bias variations.

All in all, the above paints a positive picture of the standard two-level approximation. Its regime of validity should follow from $|\epsilon(\delta\Phi^\mathrm{x})|\ll\Ec_2-\Ec_1$, where the rhs is of the order of the plasma frequency. Hence, the range of allowed $\delta\Phi^\mathrm{x}$ is actually algebraically decreasing in $g$ so that asymptotically, variations in loop current etc.\ become negligible within this range. At the same time, the range of permissible $\epsilon/\Delta$ increases exponentially, so that wide sweeps in $\epsilon$, as e.g.\ studied in the context of Landau--Zener transitions~\cite{LZ}, can be performed without leaving this range.

\subsection{Numerics}

If one numerically finds the two lowest eigenfunctions of a near-degenerate bistable rf\nobreakdash-SQUID, one can \emph{uniquely} determine the parameters of the corresponding two-level low-energy  Hamiltonian\footnote{We have used similar wordings to describe the procedure in Refs.~\cite{IMT2qb,IMT}, but the concepts are different. There, the impedance response as a function of flux bias was used to fit a single set of qubit-Hamiltonian coefficients. Here, we have numerical access to the wave functions, and can therefore determine these coefficients independently at each point in parameter space.}
\beq
  H_\mathrm{qb}=-\epsilon\sigma^z-\Delta\sigma^x\;.
\eeql{Hqb}
Namely, from the numerics one extracts the gap $\Ec_{10}\equiv\Ec_1-\Ec_0$, plus the well distribution of the ground-state wave function, $p^2=\int_{-\infty}^{\phi_\mathrm{b}}\!d\phi\,\psi_0^2(\phi)$ and $q^2=\int^\infty_{\phi_\mathrm{b}}\!d\phi\,\psi_0^2(\phi)$, where the barrier point $\phi_\mathrm{b}$ denotes the location of the potential maximum. The average energy is an irrelevant constant [assumed zero in the traceless~(\ref{Hqb})], $p$~and $q$ are constrained by normalization, and the well distribution of the first excited state is fixed by orthogonality, so that one has two pieces of data. Comparing these to the predictions of~(\ref{Hqb}), one can solve for the two unknowns as
\beq
  \epsilon=\Ec_{10}\frac{q^2{-}p^2}{2}\;,\qquad\Delta=\Ec_{10}pq\;.
\eeql{ed-res}
Thus, the coefficients of the two-state model are extracted without any handwaving, even without the artifice of an extra boundary condition stipulated below~(\ref{Eqb}); the only ingredient is a physically transparent coarse-graining of the wave functions. Note how $\epsilon$, $\Delta$ attain the expected limits at degeneracy $p=q$, and $\epsilon$ also for large bias $pq\To0$.

Another way to arrive at this result is easier to generalize to multiple qubits. It involves using the transition matrix\footnote{Sign convention: it is logical to have $\delta\Phi^\mathrm{x}<0$ correspond to $\epsilon<0$. The former favours smaller $\phi$, i.e., the left well; by (\ref{Hqb}), the latter favours $\sigma^z=-1$, so that the bottom row in $U$ corresponds to the left well. Further, the columns in $U$ and those in $H_\Ec$ of course have to be ordered consistently.} $U=\bigl(\begin{smallmatrix} q & -p \\ p & \hphantom{-}q \end{smallmatrix}\bigr)$ (i.e., having the eigenstates for its columns) to rotate the Hamiltonian $H_\Ec=\diag(\Ec_0,\Ec_1)$ from the energy basis back to the flux basis, viz.,
\beq
  UH_\Ec U^{\dagger}=\frac{\Ec_0{+}\Ec_1}{2}\bm{1}
  -\Ec_{10}\frac{q^2{-}p^2}{2}\sigma^z-\Ec_{10}pq\sigma^x\;.
\eeql{rot-H}

\begin{table}
\begin{tabular}{|c|c|c|c|} \hline
$\beta$ & $\delta\Phi^\mathrm{x}\!/\Phi_0$ & $\epsilon$ & $\Delta$ \\
\hline\hline 2 & 0 & 0 & $5.22120\cdot10^{-3}$ \\
\hline 2 &$5\cdot10^{-6}$  & $2.63\cdot10^{-5}$ & $5.22120\cdot10^{-3}$\\
%\hline 2 & $10^{-5}$  & $5.26\cdot10^{-5}$ & $5.22120\cdot10^{-3}$ \\
\hline 2 &$5\cdot10^{-5}$  & $2.63\cdot10^{-4}$ & $5.22134\cdot10^{-3}$\\
%\hline 2 & $10^{-4}$  & $5.26\cdot10^{-4}$ & $5.22178\cdot10^{-3}$ \\
\hline 2 & $5\cdot10^{-4}$  &$2.63\cdot10^{-3}$ & $5.23581\cdot10^{-3}$\\
%\hline 2 & $10^{-3}$  & $5.27\cdot10^{-3}$ & $5.27192\cdot10^{-3}$ \\
\hline 2 & $5\cdot10^{-3}$  & $2.64\cdot10^{-2}$ &$6.09330\cdot10^{-3}$\\
\hline\hline 3 & 0 & 0 & $3.00032\cdot10^{-4}$ \\
\hline 3 & $5\cdot10^{-6}$ & $2.28\cdot10^{-5}$ & $3.00032\cdot10^{-4}$\\
\hline 3 & $5\cdot10^{-5}$ & $2.28\cdot10^{-4}$ & $3.00082\cdot10^{-4}$\\
\hline 3 & $5\cdot10^{-4}$ & $2.28\cdot10^{-3}$ & $3.05012\cdot10^{-4}$\\
\hline 3 & $5\cdot10^{-3}$ & $2.28\cdot10^{-2}$ & $5.61648\cdot10^{-4}$\\
\hline\hline 6 & 0 & 0 & $5.64441\cdot10^{-6}$ \\
\hline 6 & $5\cdot10^{-7}$ & $1.38\cdot10^{-6}$ & $5.64441\cdot10^{-6}$\\
%\hline 6 & $10^{-6}$ & $2.75\cdot10^{-6}$ & $5.64441\cdot10^{-6}$ \\
\hline 6 & $5\cdot10^{-6}$ & $1.38\cdot10^{-5}$ & $5.64450\cdot10^{-6}$\\
%\hline 6 & $10^{-5}$ & $2.75\cdot10^{-5}$ & $5.64477\cdot10^{-6}$ \\
\hline 6 & $5\cdot10^{-5}$ & $1.38\cdot10^{-4}$ & $5.65360\cdot10^{-6}$\\
%\hline 6 & $10^{-4}$ & $2.75\cdot10^{-4}$ & $5.68114\cdot10^{-6}$ \\
\hline 6 & $5\cdot10^{-4}$ & $1.38\cdot10^{-3}$ & $6.50793\cdot10^{-6}$\\
\hline
\end{tabular}
\caption{Numerically extracted biases~$\epsilon$ and tunneling amplitudes~$\Delta$ according to (\ref{ed-res}), for a near-degenerate rf-SQUID with shielding~$\beta$, reduced capacitance~$g=40$, and flux bias~$\delta\Phi^\mathrm{x}$.}
\label{data}
\end{table}

Results are summarized in Table~\ref{data}. With increasing~$\beta$, one sees the same trends predicted at the end of Sec.~\ref{bist-a} for increasing~$g$: the range of validity of (\ref{Hqb}) with constant~$\Delta$ increases, in terms of not $\delta\Phi^\mathrm{x}$ but $\epsilon/\Delta$. Remaining deviations $\Delta(\delta\Phi^\mathrm{x})-\Delta(0)\sim(\delta\Phi^\mathrm{x})^2$ are too small to contradict the analytical conclusions. The breakdown of (\ref{Hqb}) occurs when the wave function in the depopulated well is so small that contributions near $\phi_\mathrm{b}$ can no longer be neglected, precluding a coarse-grained description. Even then, the linearity of $\epsilon(\delta\Phi^\mathrm{x})$ is still excellent, as shown by the last entries in Table~\ref{data} for each~$\beta$, and the breakdown is of little practical consequence as $\Delta$ is too small compared to $|\epsilon|$ to have an appreciable effect on the spectrum of~(\ref{Hqb}).

The above situation for a single qubit is uncharacteristically fortunate, in that the \emph{requirement} of unitarity on $U$ has circumvented coarse-graining the excited state~$\psi_1$. Already for two qubits, there is no natural counterpart to this shortcut, and the 4$\times$4 matrix $U$ having coarse-grained eigenstates for its columns is not exactly unitary, introducing a small uncertainty into the above procedure. The root cause of these subtleties is the problematic nature of the ubiquitous term ``flux basis,'' which strictly speaking should consist of $\delta$-functions in the coordinate representation. Pending further research, a completely satisfactory resolution may be as follows: calculate the matrix elements of $\phi$ \emph{in the low-energy subspace}. This Hermitian matrix has an eigenbasis, and the eigenvectors can be taken as columns of a transition matrix~$V$.\footnote{For consistency, the columns of $V$ should be ordered according to decreasing eigenvalues. For, say, two qubits, these columns will be the simultaneous eigenvectors of the commuting operators $\phi_a$, $\phi_b$ in the low-energy subspace.} Recognizing that $V$ consists of $\phi$-eigenvectors in the energy basis whereas $U$ consists of $\Ec$-eigenvectors in the flux basis, one should thus consider $\tilde{H}_\mathrm{qb}\equiv V^{\dagger}H_\Ec V$ as an alternative to~(\ref{rot-H}).

\section{Conclusion}
\label{discuss}

A different sign-tunable flux transformer, relying on the gradiometric nature of the employed qubit, was previously considered in~\cite{Filippov}. The transformer itself is also gradiometric, and tunability is achieved by incorporating compound junctions with variable couplings in either arm. In another type of gradiometric flux transformer, each arm of the gradiometer couples to one adjacent device, and the tunable element is a single compound junction in the central leg~\cite{Cosmelli}. This does not give sign-tunability, though: compare (4) in~\cite{Cosmelli} to (\ref{factor}) above. For the present type of mediated coupling, the prediction of sign-tunability is extremely generic: for any SQUID-type device, the $I^{(0)}(\Phi^\mathrm{x})$ curve is smooth and periodic, and therefore will have pieces with both positive and negative slope in any $\Phi_0$-interval. Comparing the two implementations, the rf-SQUID (Sec.~\ref{rf}) does not have any galvanic coupling to external circuitry, which should limit decoherence. On the other hand, it was already mentioned that a dc-SQUID coupler (Sec.~\ref{dc})~\cite{Plourde} could be useful for readout. Either seems promising, and we look forward to experimental results on tunable coupling, which may well be the next big step forward in superconducting qubit technology.

The analysis in Sec.~\ref{bist} is a prelude to the case of multiple interacting qubits, where the precise form of the low-energy Hamiltonian $H_\mathrm{qb}$ is still open to discussion. Namely, there could be multi-qubit tunneling, say, along the diagonals in a two-qubit potential, which would lead to $\sigma^x$--$\sigma^x$ and $\sigma^y$--$\sigma^y$ interactions (\cite{H2qb}, note in proof). There is little doubt that such terms would be small, but this does not automatically imply that their effect is negligible. Namely, they could, say, cause a gap at an $\mathopen|\uparrow\downarrow\rangle\leftrightarrow
\mathopen|\downarrow\uparrow\rangle$ anticrossing (for AF coupled qubits) in first order, while conventional tunneling terms $\sim\sigma^x$ contribute to this gap only in second order. Further, there seems to have been little investigation of the possibility that (single-qubit) tunneling in qubit~$a$ might depend on the state of qubit~$b$ and vice versa, which would lead to $\sigma_{a\vphantom{b}}^x\sigma_b^z$ etc.\ terms. To the extent that, for $a$-tunneling, the $b$-qubit can be regarded as an external flux bias, the flatness of $\Delta$ observed in Sec.~\ref{bist} indicates that such terms should be very small. Finally, for more than two qubits, one can anticipate three-qubit etc.\ interactions---particularly if, for stronger coupling, one qubit's persistent current can depend on the state of the others. In any case, rotation methods such as in Eq.~(\ref{rot-H}) should enable one to settle all these questions conclusively.

\section*{Acknowledgment}

We thank M.H.S. Amin, M.F.H. Steininger, and A.M. Zagoskin for fruitful discussions, and B.~Ischi and A.J. Leggett for stimulating correspondence.


\begin{thebibliography}{99}

\bibitem{mss} Y. Makhlin, G. Sch\"on, and A.~Shnirman, Rev. Mod. Phys. \textbf{73}, 357 (2001).

\bibitem{coupl-ref} Yu.A. Pashkin \emph{et al.}, Nature \textbf{421}, 823 (2003); A.J. Berkley \emph{et al.}, Science \textbf{300}, 1548 (2003).

\bibitem{IMT2qb} A. Izmalkov \emph{et~al.}, Phys. Rev. Lett. \textbf{93}, 037003 (2004).

\bibitem{feldman} X. Zhou, Z.-W. Zhou, G.-C. Guo, and M.J. Feldman, Phys. Rev. Lett. \textbf{89}, 197903 (2002).

\bibitem{AB} D.V. Averin and C. Bruder, Phys. Rev. Lett. \textbf{91}, 057003 (2003).

\bibitem{Plourde} B.L.T. Plourde \emph{et al.}, Phys. Rev. B \textbf{70}, 140501(R) (2004).

\bibitem{Mooij} E.g., J.E. Mooij \emph{et al.}, Science \textbf{285}, 1036 (1999).

\bibitem{devoret} M.H. Devoret, in \emph{Quantum Fluctuations, Lecture Notes of the 1995 Les Houches Summer School}, edited by S.~Reynaud, E.~Giacobino, and J.~Zinn-Justin (Elsevier, The Netherlands, 1997), p. 351.

\bibitem{burkard} G. Burkard, R.H. Koch, and D.P. DiVincenzo, Phys. Rev. B \textbf{69}, 064503 (2004).

\bibitem{H2qb} A.~Maassen van den Brink, Phys. Rev. B \textbf{71}, 064503 (2005).

\bibitem{IMT} M. Grajcar \emph{et~al.}, Phys. Rev. B \textbf{69}, 060501(R) (2004).

\bibitem{AFL} D.V. Averin, J.R. Friedman, and J.E. Lukens, Phys. Rev. B \textbf{62}, 11802 (2000), the appendix.

\bibitem{LZ} D.A. Garanin and R.~Schilling, Phys. Rev. B \textbf{66}, 174438 (2002); A.~Izmalkov \emph{et al.}, Europhys. Lett. \textbf{65}, 844 (2004).

\bibitem{Filippov} T.V. Filippov, S.K. Tolpygo, J.~M\"annik, and J.E. Lukens, IEEE Trans. Appl. Supercond. \textbf{13}, 1005 (2003).

\bibitem{Cosmelli} C. Cosmelli \emph{et al.}, cond-mat/0403690.

\end{thebibliography}
\end{document}